\documentclass[amsmath,amssymb,10pt,floatfix,preprintnumbers] {revtex4}
\usepackage[dvips]{graphicx}
\begin{document}
\title{ \bf Dynamics of Shock Probes in Driven Diffusive Systems}
\author{ Sakuntala Chatterjee and Mustansir Barma}
\affiliation{Department of Theoretical Physics, Tata Institute of Fundamental
Research, Homi Bhabha Road, Mumbai-400005, India}
\preprint{TIFR/TH/06-37}
\begin{abstract}
We study the dynamics of shock-tracking probe particles in driven diffusive 
systems and also in  equilibrium systems. In a driven system, they induce
a diverging timescale that marks the crossover between a passive scalar 
regime at early times and a diffusive regime at late times; a scaling form
characterises this crossover. Introduction of probes into an equilibrium 
system gives rise to a system-wide density gradient, and the presence of 
even a single probe can be felt across the entire system. 
\end{abstract}
\maketitle
A commonly used method to investigate the properties of a system is to 
 inject probe particles and monitor their behavior after they have
 equilibrated with the system under study. The 
 dynamics of probe particles then yields important information about the static
 and dynamic properties of the medium and has been used to study
 diverse phenomena, ranging from the sol-gel
 transition in a polymer solution~\cite{nakanishi} to the growth mechanism of
surfactant assemblies~\cite{hassan} and correlations present in bacterial
 motion~\cite{shiva}. The steady state involves the coupled
probe-medium system, but if the concentration of probe particles is
very low, it is generally assumed that the properties of the original
system are not affected greatly~\cite{nakanishi,hassan,shiva}. In this paper, 
we discuss the 
validity of this assumption by studying simple lattice gas models,
 and find that in certain circumstances 
 it can break down in interesting ways. We characterize
the types of behaviour that occur by studying the displacement of a probe
particle as a function of time, the correlations between different
probes particles, and their effect on the medium. Depending on the
system under study, different behaviours are found in
the low-concentration limit, ranging from diverging correlation
lengths and power law decays, to effects which are felt over
macroscopic distances throughout the system. We relate these
differences to an interesting interplay between the equilibrium and
nonequilibrium characteristics of the medium and the probe particles.

We focus on the effects of probe particles on one-dimensional driven
diffusive systems, and their equilibrium counterparts. We mainly consider
probes which exchange with the particles and holes of the medium with equal
rate but in opposite directions. Such probes  
tend to settle in regions of strong
density variations (shocks) and thus serve to track the locations and
movements of shocks. We consider primarily, but not exclusively,
a medium in which the shock-tracking probes
(STPs) reduce to second class particles~\cite{fer3,derrida}, {\sl i.e.} they
 behave as
holes for the particles and as particles for the holes. 
STPs are intrinsically nonequilibrium
probes, as they evolve through moves which do not respect detailed
balance. We will see below that they have very strong effects on a
medium that is originally in equilibrium: even a single probe
generates a density perturbation which extends over macroscopic
distances. However, when immersed in a medium that is in a
current-carrying nonequilibrium state, the perturbation produced by a
single probe particle decays to zero, but as a slow power law;
 a finite density of probes brings in a correlation length which diverges
strongly in the low concentration limit. We find that these broad
conclusions remain valid for a class of models with nearest neighbor
interaction and also with an extended range
of particle hopping. 

 Recently, Levine {\sl et al.}~\cite{lev,lev2} have
 studied the behavior of STPs in a driven diffusive system whose steady state
 has nearest neighbor Ising measure~\cite{kls}. When only a finite number of
 probes are present, 
 the medium induces a long-ranged  attraction between
 pairs of successive 
 STPs, as shown earlier by Derrida {\sl et al.}~\cite{derrida}.
 The distribution of the separation between a  
pair follows a power law with a power which
depends continuously on the strength of the Ising interaction~\cite{lev}.
Consequences for the phase diagram
with a macroscopic number of STPs were explored in~\cite{lev2} . In this
paper, we are  interested in the {\it dynamical} properties
of STPs primarily for the case of uncorrelated particle occupancies. We find
that these properties are governed by a time-scale $\tau$ which
diverges as the STP concentration approaches zero. This time-scale
is related to a diverging correlation length which characterises the density
profile around a probe particle~\cite{derrida}. $\tau$ enters into a scaling
description of the crossover between a `passive scalar advection' regime
and a diffusive regime which sets 
in at longer times. Similar behaviour is found, though with some
interesting variations, also in members of the class of $k$-hop
models, which are driven diffusive systems with longer ranged hopping
of particles~\cite{bind}.

Even more striking is the effect of STPs on a
system which is initially in equilibrium. With the introduction of a
single probe, the medium develops a density gradient across
macroscopic distances and a concomitant shock around the probe. Interestingly,
the resulting state shows a signature of  an inhomogeneous product
measure. The  dynamical properties of the medium 
remain characteristic of local
equilibrium, despite the fact that there is a small current in the system. 

The system is defined on a 
$1$-d lattice with periodic boundary conditions. 
We denote particles and holes in the original system by `$+$' and `$-$',
respectively, and the probe particles by `$0$'. The dynamical
moves involve the exchanges~\cite{fer3,derrida}:  
\begin{eqnarray}
\nonumber
(a) \;\;\;\; +- &\rightarrow& -+\\
\label{eq:rate}
(b) \;\;\;\;\; +0 &\rightarrow& 0+\\
\nonumber
(c) \;\;\;\;\;\;\; 0- &\rightarrow& -0
\end{eqnarray}
All moves take place with equal rate. Since a probe exchanges with a particle
and a hole in opposite directions, it tends to migrate to places where
there is an excess of holes to its left and particles to its right, {\sl i.e.}
a shock.  

In the absence of STPs the system reduces to the asymmetric simple
exclusion process (ASEP), which is a paradigm of driven diffusive
systems~\cite{lig}. 
With probes present, we see that
a particle (hole) exchanges with a hole (particle) and a probe in
exactly the same way. The STPs
in this case reduce to second class particles~\cite{fer3,derrida}. If
$\rho$ and $\rho_0$ are the densities of particles and probes
respectively, then a particle behaves as
if in an ASEP with an effective hole density $(1-\rho)$, while a hole
finds itself in an ASEP with an effective particle density $(\rho +\rho_0)$.
STPs are correlated but the exact steady state measure 
for the system can be found using the matrix method~\cite{derrida}. 
Throughout this paper, we will consider equal densities of particles
and holes in the medium, which implies $\rho_0 = 1-2 \rho$.

\begin{figure}
\includegraphics[angle=0,scale=0.7]{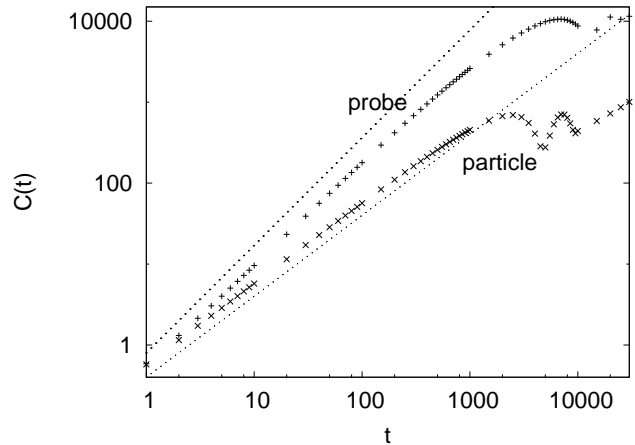}
\caption{\it Mean squared displacement of particles and probes for 
$L=2048$ and $\rho_0=0.15$, showing the initial linear growth of $C_+(t)$ and
the nonlinear growth of $C_0(t)$. The dashed lines depict power laws with
exponent $4/3$ and $1$. 
The late-time oscillations are due to the existence of kinematic
waves in a finite periodic system.}
\label{fig:kw}
\end{figure}

To monitor the dynamics,
we measured the mean squared displacement of tagged particles
in the medium, and tagged STPs, using Monte
Carlo simulation after the system has reached steady state.
Let $Y_k(t)$ be the position of the $k$th particle at 
time $t$. We monitor the mean 
squared displacement defined as 
\begin{equation}
C(t) = \langle \left ( 
Y_k (t)-Y_k (0)- \langle Y_k (t)-Y_k (0) \rangle \right )^2 \rangle
\end{equation}
where the average is over the steady state ensemble of 
initial configurations. We denote by $C_0(t)$ and $C_+(t)$ the mean squared
displacements of probes and particles, respectively. Monte Carlo simulation  
results show interesting differences between $C_+(t)$ and $C_0(t)$ with a low
but finite density of probes [Fig.(\ref{fig:kw})]. In contrast to the 
linear increase of $C_+(t)$,
$C_0(t)$ increases non-linearly in the range of time considered. The behaviors
of both $C_+$ and $C_0$, including long-time crossovers, 
are discussed in detail below.

Tagged particles in the medium behave as if in a regular ASEP, 
for which $C_+(t)$ is known to  grow linearly in
time for an infinite system~\cite{fer2}. 
In a finite system of size $L$, $C_+(t)$ is nonmonotonic due to
the existence of a kinematic wave which carries
density fluctuations through the system with speed
$\frac{dJ}{d\rho} =(1-2\rho )$~\cite{light,schutz},
where $J$ is the current, given by
$\rho (1-\rho )$. Since the average speed of
the tagged particle is $J/\rho =(1-\rho )$, it moves from one density 
patch to the
other with relative speed $\rho$; the
 mean squared displacement increases linearly, since each
 patch contributes a random excess to the relative velocity of the tagged
 particle. The subsequent dips
in the mean squared displacement, as seen in Fig.(\ref{fig:kw}),
correspond to the return of a tagged particle
to its initial environment in time $L/\rho$; the lower envelope of the
curve in Fig.(\ref{fig:kw}) grows as $t^{2/3}$ as long as
$t \ll L^z$~\cite{maya,snm}. 

\begin{figure}
\includegraphics[scale=0.7,angle=0]{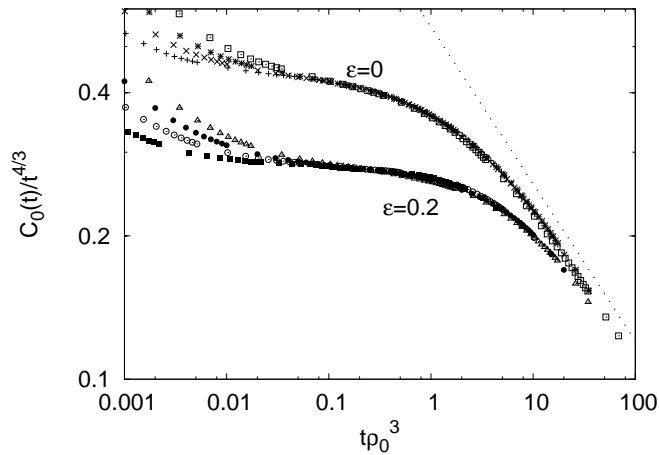}
\caption{\it Scaling collapse for mean squared displacement of tagged 
probe particles with densities  $0.08,0.1,0.12,0.15$ $($moving upwards$)$. 
The upper curve shows $C_0 (t)$ for $\epsilon =0$ $(L=131072$, averaged 
over $100$ histories$)$ and the lower curve shows $C_0 (t)$ for 
  $\epsilon =0.2$ $(L=16384$, averaged over $1000$ histories$)$. The dashed
 line shows a power law decay with exponent $-1/3$. } 
\label{fig:fertag}
\end{figure}

 For tagged probe particles, $C_0(t)$ shows a crossover from an initial 
 passive scalar advection regime to a long-time 
diffusive regime. The associated crossover occurs on 
a timescale $\tau$ that diverges strongly as the
probe density approaches zero.  
Ferrari and Fontes have calculated the
asymptotic ($t \rightarrow \infty$) 
behavior of $C_0(t)$ for STPs~\cite{fer} 
and shown that  $C_0(t) \approx Dt$ with diffusion constant
$D=\left [ \rho(1-\rho) + (\rho + \rho_0 )(1-\rho - \rho_0) \right ]/\rho_0 $. 
For small times,  in the limit of low concentration of 
the probe particles, one would expect that each STP would
behave as an individual non-interacting
particle, subject only to the fluctuations of the medium. The mean squared
displacement of a single probe has been shown analytically to grow as $t^{4/3}$
using the matrix product method~\cite{kirone}.
 This behavior agrees with the results for particles sliding down on surface 
 evolving through Kardar-Parisi-Zhang dynamics,  where $C_0 (t)$
 grows as $t^{2/z}$, with
$z=3/2$~\cite{nagar}. We propose 
the following scaling form in the limit
of large $t$ and $\tau$ 
\begin{equation}
C_0(t)  \sim t^{4/3} F \left ( \frac{t}{\tau}  \right ) 
\label{eq:tag}
\end{equation}
in terms of the crossover time $\tau$ whose dependence on $\rho_0$ is given
below.
Here $F(y)$ is a scaling function which approaches a constant as 
$y \rightarrow 0$.
For $y \gg 1$, we must have  $F(y) \sim y^{-1/3}$, in order to reproduce
$C(t) \approx Dt$. The timescale $\tau$ is related to 
the correlation length: at sufficiently large distance $r$ from an
STP, the shock profile decays as $r^{-1/2}\exp(-r/\xi)$ where
$\xi$ has been calculated by Derrida {\sl et al.}~\cite{derrida}.
In the limit of small $\rho_0$, it reduces to  
$ \xi \approx 4\rho(1-\rho)/\rho_0^2$ which diverges as  
$\rho_0 \rightarrow 0$. Relating $\tau$ to $\xi$  
through  $\tau \sim \xi^z$ with $z=3/2$ then implies $\tau \sim \rho_0 ^{-3}$. 
We verified the scaling form by Monte Carlo simulation. We were able to
circumvent equilibrating by directly generating  
steady state initial configurations, 
following the prescription in~\cite{fap}.   
In Fig.(\ref{fig:fertag}), we  plot $C(t)/t^{4/3}$ vs $t/\tau$ for 
various values of $\rho_0$  and obtain a good scaling collapse, except for
very small values of $t$ which fall outside the scaling regime.

A nontrivial
check of the scaling form comes from examining the dependence of $\tau$ 
on $\rho_0$. Matching the early and late time form  for $C_0(t)$  at  
$\tau \sim \rho_0 ^{-3}$ then yields $D \sim \rho_0^{-1}$,
in agreement with~\cite{fer}. Yet another check comes from
considering the implication for a system with a finite size $L$. Finite size
scaling would suggest that once $L$ is smaller than $\rho_0 ^{-2}$, the
behavior $D \sim \xi^{1/2}$ found above should give way to $D \sim L^{1/2}$.
This is in conformity with the calculation of~\cite{evans}.

\begin{figure}
\includegraphics[scale=0.7,angle=0]{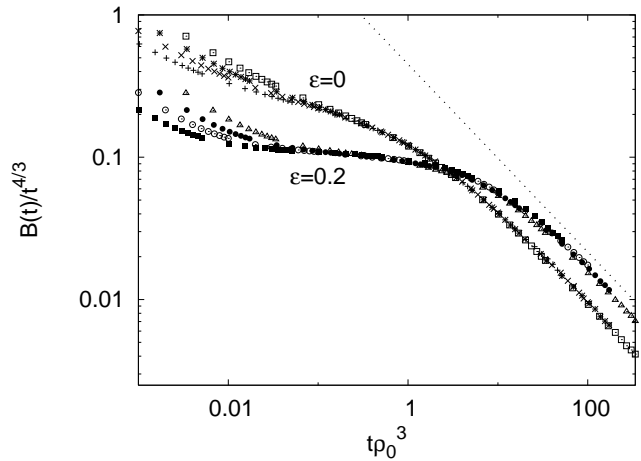}
\caption{\it Scaling collapse for $B(t)$ with probe densities
$0.08,0.1,0.12,0.15$ (moving upawrds). The two curves correspond to 
$\epsilon=0$ and $\epsilon=0.2$  ($L=16384$,
averaged over $25$ initial configurations and $40$ evolution histories for
each). The dashed line shows
a power law decay with exponent $-2/3$.}
\label{fig:van}
\end{figure}

To track dissipation in a single component system, one can perform a Galilean
shift to keep up with the kinematic wave~\cite{maya}. This is not possible in
our system since there are two kinematic waves with different velocities
$\pm (1-2\rho)$ corresponding to the two ASEPs with densities $\rho$ and 
$\rho + \rho_0$. 
We thus use the method of van Beijeren~\cite{van}, wherein
we monitor the quantity 
\begin{equation}
B(t)=\overline{\left ( Y_k(t)-Y_k(0)-\overline{(Y_k(t)-Y_k(0))}\right )^2}
\end{equation}
where the overhead bar denotes averaging over different evolution
histories, starting from a {\it fixed} initial configuration drawn from 
the steady state ensemble. Note that in this
averaging process, the initial pattern of density fluctuations
around a particular tagged probe is identical for all evolution histories.
The mean $\overline{(Y_k(t)-Y_k(0))}$ shows fluctuations superposed on a
linear growth law. These fluctuations are determined by the density pattern in
the initial configuration~\cite{shamik}.   
 $B(t)$ therefore gives the spread of this pattern with time. 
 We find that $B(t)$ increases as $t^{2/3}$ for large $t$, and  would expect a
 scaling function to connect this regime to the small time regime 
 $B(t) \sim t^{4/3} $, characteristic of single particle behavior. In 
 the limit of large $t$ and small $\rho_0$, we expect 
 \begin{equation}
 B(t) \sim t^{4/3} G\left ( \frac{t}{\tau} \right ). 
\label{eq:van}
\end{equation}
The scaling function $G(y)$
should approach a constant as $y \rightarrow 0$ while for $y \gg 1$, one
expects 
$G(y) \sim y^{-2/3}$. Our simulation results [Fig.(\ref{fig:van})] are
consistent with this scaling form. 
 
 Similar
results are found for shock trackers in a driven diffusive system in which
there is an Ising interaction 
$V=-\epsilon [(n(i)-\frac{1}{2})(n(j)-\frac{1}{2})]$
between the neighboring particles $\langle ij \rangle$
 in the medium~\cite{kls}. The
dynamical moves are as specified in Eq.(\ref{eq:rate}) but the rates are  
different: while the moves $(b)$ and $(c)$ occur with rate $1$, move $(a)$ 
takes place with a rate $(1-\Delta V)$ where 
$\Delta V$ is the change in Ising energy~\cite{kls}. Note that in this 
case, STPs do not behave as second class particles. 
In~\cite{lev}, it has been shown that with a finite number
of probes, the medium induces
an attraction and 
the separation $s$ between two consecutive probes follows the
distribution $P(s) \sim s ^{-b}$ for large $s$. The coefficient $b$ varies 
continuously with $\epsilon$; in the
non-interacting case, $b=3/2$. If $s_i$ is the separation between the
$i$-th and
$(i+1)$-th probe and $R_m$ is the distance between the first and the $(m+1)$-th
probe, then $R_m=\sum_{i=1}^{m} s_i$. The quantity $R_m$ then follows a 
 L\'evy distribution with a norming constant 
$\sim m^{1/(b-1)}$, so long as $R_m < \xi$.
This in turn implies that the correlation length  
$\xi \sim \rho_0 ^{-1/(2-b)}$ and hence $\tau \sim \xi ^{z_0} \sim \rho_0
^{-z_0/(2-b)}$, where $z_0$ is the dynamical critical exponent of the system.
Our numerical simulations 
indicate that for the ranges of $\epsilon$,
$\rho_0$ and $t$ studied (Fig.(\ref{fig:fertag})and (\ref{fig:van}) ), 
$C_0(t)$ and $B(t)$ continue to show a crossover at a timescale 
$\tau \sim \rho_0 ^{-3} $, surprisingly similar to the 
non-interacting case~\cite{sc}.

To investigate the effect of probes in an equilibrium system, we
studied STPs in a symmetric exclusion
process (SEP). The dynamical rules remain the same as in Eq.(\ref{eq:rate}),
except that $(a)$ changes to $+- \rightleftharpoons -+$. All 
moves take place with equal rate.
This model is a special case  of the model proposed by Arndt
{\sl et al.} with the asymmetry parameter set equal to its critical value
$q=1$~\cite{ahr}.  

In the absence of probes the system obeys detailed balance and the state is
described by a uniform product measure. With the introduction of a
single probe, the condition of detailed
balance is violated and there is a small ($\sim 1/L$) current in the system.
 In this nonequilibrium steady state, there is a system-wide density
gradient around the probe.

\begin{figure}
\includegraphics[scale=0.7,angle=0]{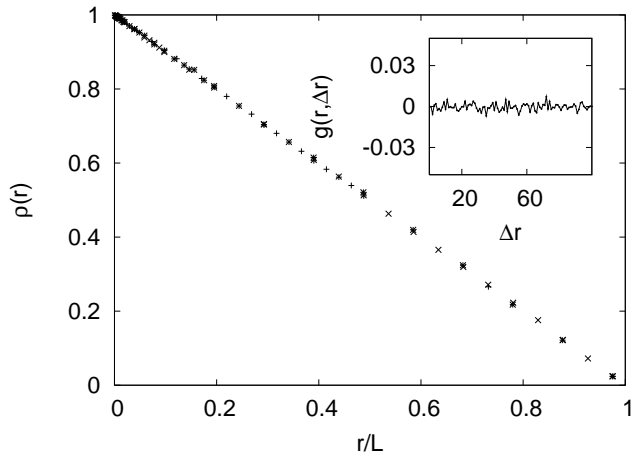}
\caption{\it Density profile as a function of the scaled distance away 
from the probe in the SEP for $L=513,1025,2049$. 
 The data is averaged over $50000$ histories. The inset shows 
 $g(r,\Delta r)$ for $r =1024$ $(L=2049$, 
 averaged over $10^4$ histories$)$ and illustrates that the pair correlation
 is close to zero.}
\label{fig:rho}
\end{figure}

Figure (\ref{fig:rho}) shows the density  
$\rho(r) \equiv \langle n(r) \rangle$ , at a distance $r$ away from the probe,
where $n(r)$ is the occupancy at $r$. 
 We find that $\rho(r)=A(1-r/L)$ with $A \simeq 1$.
The inset in Fig.(\ref{fig:rho}) shows the spatial two point correlation
function defined as
$g(r,\Delta r)=\langle n(r) n(r+\Delta r) \rangle - \rho(r)\rho(r+\Delta r)$, 
for a fixed value of $r$.  
The fact that $g(r,\Delta r)$  is close to zero points
to the existence of an inhomogeneous product measure state.   

\begin{figure}
\includegraphics[scale=0.7,angle=0]{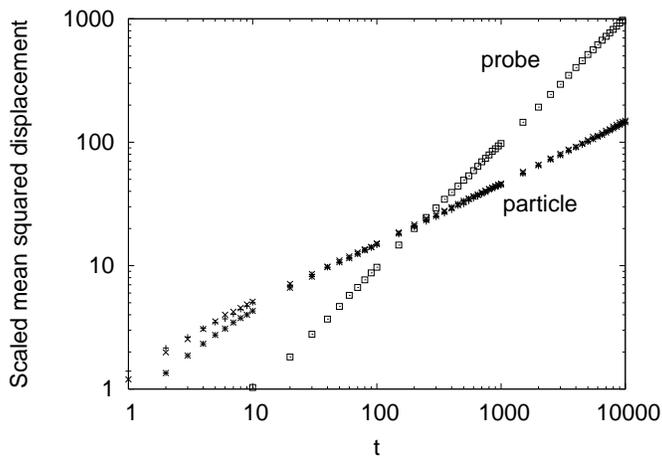}
\caption{\it Scaled mean squared displacement of tagged particles at distances
$r=580,985,1390$, away from the probe, 
corresponding to  $\rho (r)=0.72,0.52,0.32$. $(L=2049$, averaged
over $10^4$ histories$)$. The curves are seen to merge when the coefficient 
${\left ( \frac{2}{\pi} \right )}^{1/2} \frac{1-\rho (r) }{\rho (r)}$
 is divided out. Also shown is the mean squared displacement
 of a single probe scaled up by a factor of 100.}
\label{fig:tagsep}
\end{figure}

The mean squared displacement of the probe grows diffusively with a diffusion
constant $D \sim 1/L$ (Fig.(\ref{fig:tagsep})).
This follows from the fact that for equal densities of
particles and holes in the medium, the probe has  an equal probability to move
to the left or to the right and from the form of $\rho(r)$ given above, it
follows that this probability $\sim 1/L$ in each Monte Carlo step. 
 Far from the probe, the local 
properties of the medium still resemble those of the SEP. We demonstrate this
by measuring the mean squared displacement of a tagged particle $C_+(r,t)$ 
and comparing with the SEP result  
$C_+(t) \approx \sqrt {\left ( \frac{2}{\pi} \right )} (1-\rho )/\rho
\mbox{ } t^{1/2} $~\cite{lig}. In Fig.(\ref{fig:tagsep}), we present data for
$C_+(r,t)$ for different values of local densities.

 This agreement should hold as long as the tagged
particle remains in the region where the local density is $\rho(r)$. Since the
density changes over a length scale $\sim L$ and the average velocity of the
tagged particle is $\sim 1/L$, the region of validity extends upto a 
time scale  $\sim L^2$.  

When a macroscopic number of STPs is introduced, they are phase 
separated and form essentially 
a single cluster. The  other phase is comprised of the  
medium, which continues to remain in local
equilibrium with a similar density gradient as in the single probe case. The
probe cluster is found to diffuse slowly through the medium with a diffusion 
constant $D \sim 1/L$.

So far we have considered the effect of shock-tracking probes on systems 
 described by the simple exclusion models and the model with Ising interactions.
Our broad
 conclusions from these studies remain valid even for systems with an
 extended range for particle hops. We have studied the behavior of STPs in a 
 driven diffusive system described by the $k$-hop model, in which the
 particle hopping range is $k$~\cite{bind}. 
  Our simulation shows that in this case also, the
effect produced by the probe is long-ranged---there is a diverging
correlation length in the low concentration limit of 
STPs~\cite{sc}. In the symmetric version of the $k$-hop model,  
a single STP gives rise to a macroscopic density gradient, as before.

In this paper, we have discussed the effect of shock-tracking probes on 
one dimensional 
equilibrium and nonequilibrium systems. An equilibrium system is found to be
affected very strongly by the presence of even a single probe---a density
gradient is generated that extends over the entire system. For probes in 
nonequilibrium systems, the effect is less drastic: the shock
around a single probe decays as a slow power law. With a macroscopic number
of probes, this decay extends upto the correlation length, which in turn 
diverges in the low probe-density limit. The question arises: How general are
these conclusions---do they remain valid even with other nonequilibrium
probes in
one-dimensional systems? We have studied another example which falls into
this general framework, namely directed probes driven by an external field. 
In an ASEP, such a probe behaves like a tagged 
particle, but when immersed in a symmetric exclusion process, 
it is again found to give rise to a macroscopic
density gradient~\cite{sc}. It would be interesting to have a general
criterion which sets conditions for the occurrence of such phenomena. 
 
{\bf Acknowledgements:} We acknowledge useful discussions with P.A. Ferrari, 
G.M. Sch\"{u}tz and D. Mukamel.
SC would like to thank the TIFR Endowment Fund for
partial financial support. MB acknowledges the hospitality of the Isaac Newton
Institute, Cambridge and support through ESPRC grant
$531174$.



\begin{thebibliography}{99}
\bibitem{nakanishi} H. Oikawa and H. Nakanishi, {\it J. Chem. Phys.} {\bf 115},
3785 (2001).
\bibitem{hassan} P.A. Hassan and C. Manohar, {\it J. Phys. Chem. B} {\bf 102}, 
7120 (1998).
\bibitem{shiva} G.V. Soni, B.M. Jaffar Ali, Y. Hatwalne and G.V. Shivashankar
{\it Biophys. J.} {\bf 84}, 2634 (2003).
\bibitem{fer3} P.A. Ferrari, C. Kipnis and S. Saada {\it Ann. Prob.} 
{\bf 19}, 226 (1991).
\bibitem{derrida} B. Derrida, S.A. Janowsky, J.L. Lebowitz and E.R. Speer 
{\it J.Stat. Phys.} {\bf 73}, 813 (1993).
\bibitem{lev} E. Levine, D. Mukamel and G.M. Sch\"{u}tz {\it Europhys. Lett.}
{\bf 70}, 565 (2005).
\bibitem{lev2} Y. Kafri, E. Levine, D. Mukamel, G.M. Sch\"{u}tz and 
R.D. Willmann
{\it Phys. Rev. E} {\bf 68}, 035101 (R) (2003);
C. Godr\`eche, E. Levine and D. Mukamel
{\it J. Phys. A: Math. Gen}  {\bf 38}, L523 (2005); M.R. Evans, E. Levine, 
P.K. Mohanty and D. Mukamel
{\it Eur. Phys. J. B} {\bf 41}, 223 (2004).
\bibitem{kls} S. Katz, J.L. Lebowitz and H. Spohn {\it J. Stat. Phys.}
{\bf 34}, 497 (1984).
\bibitem{bind} P.M. Binder, M. Paczuski and M. Barma {\it Phys. Rev. E}
{\bf 49}, 1174 (1994)
\bibitem{lig} T.M. Liggett {\it Interacting Particle Systems}
(Springer-Verlag) (1985).
\bibitem{fer2}A. De Masi and P.A. Ferrari {\it J. Stat. Phys.} {\bf 38}, 667
(1984).
\bibitem{light} M.J. Lighthill and G.B. Whitham {\it Proc. R. Soc. London A}
{\bf 229}, 281 (1955).
\bibitem{schutz} G.M. Sch\"{u}tz in {\it Phase Transitions and Critical 
Phenomena }, vol. 19, edited by C. Domb, J.L. Lebowitz (Academic, London, 2001).
\bibitem{maya} M. Paczuski, M. Barma, S.N. Majumdar and T. Hwa 
{\it Phys. Rev. Lett.} {\bf 69}, 2735 (1992).
\bibitem{snm} S.N. Majumdar and M. Barma {\it Phys. Rev. B} {\bf 44}, 5306 (1991).
\bibitem{fer} P.A. Ferrari and L.R.G. Fontes {\it Prob. Theory Related Fields} 
{\bf 99}, 305 (1994).
\bibitem{kirone} C. Boutillier, P. Fran\c{c}ois, K. Mallick, and S. Mallick 
{\it J. Phys. A: Math. Gen.} {\bf 35}, 9703 (2002). 
\bibitem{nagar} A. Nagar, S.N. Majumdar and M. Barma {\it Phys. Rev. E} 
{\bf 74}, 0211124 (2006).
\bibitem{fap} O. Angel {\it J. Comb. Theory, Ser.A} {\bf 113}, 625 (2006).
\bibitem{evans} B. Derrida and M.R. Evans {\it J. Phys. A: Math. Gen.} {\bf
32}, 4833 (1999).
\bibitem{van} H.van Beijeren {\it J. Stat. Phys.} {\bf 63}, 47 (1991).
\bibitem{shamik} S. Gupta, S.N. Majumdar, C. Godr\`eche and M. Barma, in
preparation.
\bibitem{sc} S. Chatterjee and M. Barma,  in preparation.
\bibitem{ahr} P.F. Arndt, T. Heinzel and V. Rittenberg {\it J. Phys. A} 
{\bf 31}, L45 (1998); {\it J. Stat. Phys.} {\bf 97}, 1 (1999). 
\end{thebibliography}
\end{document}